\documentclass[10pt,conference]{IEEEtran}
\IEEEoverridecommandlockouts
\usepackage{cite}
\usepackage{multirow}
\usepackage{graphicx}
\usepackage{rotating}
\usepackage[export]{adjustbox}
\usepackage{amsmath,amssymb,amsfonts}
\usepackage{algorithmic}
\usepackage{graphicx}
\usepackage{afterpage}
\usepackage{textcomp}
\usepackage{pdfpages}
\usepackage{hyperref}
\usepackage{caption}
\usepackage{subcaption}
\usepackage{xcolor}
\usepackage{placeins}
\usepackage{multicol}
\usepackage{url}
\usepackage{lipsum}  
\usepackage{verbatim}
\usepackage[inline]{enumitem}
\usepackage{setspace}
\usepackage{float}
\usepackage{tabularx}
\usepackage{adjustbox}
\usepackage{array}
\usepackage{enumitem}   
\usepackage{array}
\usepackage{multirow}

\newcolumntype{P}[1]{>{\centering\arraybackslash}p{#1}}
\renewcommand\thesubsection{\Alph{subsection}}

\hypersetup{
    pdfborderstyle={/W 0.5}, 
}

\begin{document}

\title{Improving Web Content Delivery with HTTP/3 and Non-Incremental EPS}

\author{\IEEEauthorblockN{\textsuperscript{} Abhinav Gupta and Radim Bartos}
\IEEEauthorblockA{\textit{Department of Computer Science} \\
\textit{University of New Hampshire}\\
Durham, NH 03824, USA \\
{\{ag1226,rbartos\}}@cs.unh.edu}}

\maketitle

\begin{abstract}

HTTP/3 marks a significant advancement in protocol development, utilizing QUIC as its underlying transport layer to exploit multiplexing capabilities and minimize head-of-line blocking. The introduction of the Extensible Prioritization Scheme (EPS) offers a signaling mechanism for controlling the order of resource delivery. In this study, we propose mappings from Chromium priority hints to EPS urgency levels with the goal of enhancing the key web performance metrics. The mappings are evaluated using EPS's urgency-based, non-incremental resource delivery method. The results of the experimental evaluation show that the proposed mappings improve the Quality of Experience metrics across a range of websites.

\end{abstract}

\begin{IEEEkeywords}
HTTP/3 Prioritization, Extensible Prioritization Scheme, Quality of Experience, Lighthouse, Protocol Performance 
\end{IEEEkeywords}

\section{Introduction}

The timely delivery of web resources is pivotal in enhancing website users' Quality of Experience (QoE). In this context, HTTP/3~\cite{rfc9114} represents a significant advancement as the latest protocol, utilizing QUIC~\cite{rfc9000} as its underlying transport layer protocol. HTTP/3 leverages the multiplexing capabilities of QUIC to efficiently transfer data, significantly positioning it as the successor to HTTP/2. QUIC outperforms TCP in terms of connection setup speed and is capable of achieving a zero RTT (Round-Trip Time) connection setup for connections that have been established previously. Moreover, it addresses and mitigates the Head-of-Line (HoL) blocking issue prevalent with HTTP/2. 

To further enhance the timely delivery of resources, we utilize the Extensible Prioritization Scheme (EPS)~\cite{rfc9218}, which serves as a successor to the more complex prioritization strategies of HTTP/2. This evolution aims to streamline and optimize the process of resource delivery, addressing the complexity previously encountered with HTTP/2's prioritization mechanisms~\cite{http2_prioritization}.

EPS~\cite{rfc9218} describes two principal strategies for transmitting web resources: incremental delivery, analogous to concurrent delivery, and non-incremental delivery, akin to sequential delivery. Each approach significantly contributes to the improvement of QoE, underscoring the importance of choosing the appropriate delivery method based on the specific needs and context of the website.

In this study, we use non-incremental scheduling, where transmission simplifies the process by sending resources one after the other. This straightforward approach can significantly reduce the complexity on the server side and mitigate issues related to resource contention and dependency resolution. Moreover, non-incremental transmission ensures a predictable order of delivery, which is crucial for resources with interdependencies, as it guarantees that prerequisites are loaded before dependent resources.

This paper explores the impact of prioritization on the QoE by employing an HTTP/3-based \emph{aioquic}~\cite{aioquic} server, augmented with the EPS, within a controlled testbed. Key to our investigation is utilizing multiple prioritization mapping strategies, which we identify as instrumental in improving the performance. 

To evaluate the impact of various prioritization mappings, we utilize the industry-standard tool, Lighthouse~\cite{lighthouse}, for our tests. Our assessment focuses on six key performance metrics~\cite{lighthouse_metric}: First Contentful Paint (FCP), Largest Contentful Paint (LCP), Time to Interactive (TTI), Total Blocking Time (TBT), Speed Index (SI), and Cumulative Layout Shift (CLS). These metrics collectively provide a comprehensive view of web performance and user experience.

The main contributions of this paper are as follows:
\begin{itemize}
    \item We augment a standard HTTP/3-based \emph{aioquic} server with the urgency-based non-incremental resource delivery method specified by the EPS within a controlled test environment.
    \item We propose two prioritization mapping strategies and we investigate their effects on the QoE across eight widely used websites, comprehensively analyzing their impacts.
    \item Our findings demonstrate that adopting an EPS urgency-based non-incremental resource delivery method improves QoE compared to the default sequential scheduling provided by the standard \emph{aioquic} server.
\end{itemize}

The rest of the paper is organized as follows: Section~\ref{background} provides the background for prioritization and HTTP/3. Section~\ref{related_work} reviews related work on prioritization, HTTP/3, and QUIC performance. Section~\ref{proposed_mapping} details our proposed mapping strategies. Section~\ref{experimental_setup} describes the experimental setup, and Section~\ref{experimental_evaluation} analyzes the results achieved from our proposed mappings. The paper concludes in Section~\ref{conclusions} with a summary of our findings and directions for future research.

\section{Background}
\label{background}
The HTTP has been a foundational element of web communication, reflecting and adapting to the evolving demands of internet use. HTTP/2 marked a significant upgrade from its predecessors by introducing multiplexing, which allowed multiple resource streams to be transported concurrently over a single TCP connection.

However, the reliance on TCP introduced its own set of challenges. Notably, TCP's susceptibility to HoL blocking became a critical concern, where the loss of a single packet could stall the entire data stream. Additionally, TCP's connection establishment time can reach up to 3RTTs when SSL authentication is involved, reducing the speed and efficiency of web interactions~\cite{QUICTCPPerfromance}.  

HTTP/2's prioritization mechanism, intended to optimize web loading times by allowing browsers to signal their resource preferences, added another layer of complexity. The use of a weight-based dependency tree to represent resource priorities was conceptually sound but proved challenging to implement effectively\cite{rmarx_resource}. The servers often ignored the browsers' prioritization signals or implemented them incorrectly, leading to suboptimal loading strategies and degraded user experiences~\cite{http2_prioritization_issues}. 

These operational inefficiencies underscored the necessity for a transport protocol that could surmount the limitations of TCP. QUIC emerged as a promising alternative designed to tackle the shortcomings of TCP by facilitating multiple independent streams and minimizing HoL blocking. It leveraged the speed and simplicity of UDP while incorporating essential features of TCP, such as reliable delivery and congestion control, to ensure robust performance even in challenging network conditions\cite{rfc9000}.

The maturation of QUIC catalyzed the development of HTTP/3, which was standardized in June 2022. HTTP/3 aligns with QUIC's strengths, offering reduced connection establishment times, improved handling of packet loss, and the ability to maintain connectivity even during IP address changes, a common occurrence in mobile environments.

Concurrently, the EPS\cite{rfc9218} was introduced, representing a paradigm shift in resource prioritization. Standardized alongside HTTP/3 in June 2022, EPS supplanted the complex dependency tree with a more manageable system of stream urgencies and incremental flags. The scheme, as detailed by Cloudflare~\cite{cloudflare2023http3}, represents a significant departure from previous prioritization mechanisms~\cite{http2_prioritization}, offering a more adaptable and efficient approach to resource delivery.  

This model conferred the flexibility to send streams with different EPS urgency levels in either an incremental or a non-incremental manner, which not only simplifies the process of prioritization implementation but is also particularly pertinent for the rendering of modern web pages, where the sequencing of resource delivery is critical.

Resources like JavaScript and CSS, pivotal for page structure and appearance, can be render-blocking, requiring strategic prioritization. JavaScript, in particular, can alter the HTML document's structure dynamically, necessitating its early retrieval and processing\cite{rmarx_resource}. 

The updated approach to prioritizing resources and the insights from web performance metrics lead us to a review of related work in the domain.

\section{Related Work}
\label{related_work}

While the development and implementation of efficient scheduling mechanisms within web protocols have been extensively explored over a long period, the transition from HTTP/2 to HTTP/3, alongside the standardization of the EPS~\cite{rfc9218} and QUIC~\cite{rfc9000}, marks a new frontier in this domain. This evolution highlights the shift in performance and prioritization studies, opening up many uncharted research avenues. These advancements underscore the potential for innovation and further exploration in improving web performance.

A study on SPDY's performance~\cite{how_spdy_is_spdy} found significant improvements over HTTP through its single TCP connection. However, these advantages can be mitigated by page dependencies and browser computations, leading to modest gains in specific scenarios. In a separate effort, KLOTSKI~\cite{Klotski} was introduced to enhance the mobile web user experience by prioritizing content based on user preferences, effectively addressing inter-resource dependencies without necessitating website modifications. It was found that KLOTSKI significantly improved the user experience compared to native websites. Furthermore, research on Polaris~\cite{netravali} showcased a dynamic client-side scheduler that leverages fine-grained dependency graphs to significantly reduce page load times by optimizing resource fetching sequences. Polaris reduced median page loading times by 34\% and achieved a 59\% reduction at the 95\textsuperscript{th} percentile. 

A study by Hugues de Saxcé et al. on HTTP/2~\cite{http2_faster_http1}, addressing its evolution from HTTP/1.1, delves into features like compression, multiplexing, server push, and priority. It highlights that while these advancements aim to rectify past protocol inefficiencies and are poised to improve web performance, the reality of packet loss in cellular networks can dampen these benefits.

Building on the foundational insights into HTTP/2's capabilities, subsequent research further explores the realm of prioritization within this framework. A detailed study by Wijnants et al.~\cite{h2_wijnants} involving eight prioritization algorithms, two user agents, 40 realistic webpages, and diverse network conditions reveals that while complex prioritization strategies can markedly improve page load times, simpler, naive approaches may result in visual load times slowing by over 25\%.

Moving forward with QUIC, the protocol enhances connection speed and mitigates head-of-line blocking in unstable networks, outperforming even optimized TCP configurations~\cite{performance_perspective_quic}. QUIC outperforms under poor network conditions due to its lower latency and enhanced congestion control, though its advantages diminish with webpages composed of numerous small objects~\cite{WebQUICFaster}. Further research supports QUIC's effectiveness over HTTP and SPDY, particularly in lossy environments, showcasing its potential to advance web communication efficiency~\cite{HTTPoverUDP,QUICBetterforWhom}. Google's extensive deployment of QUIC, underpinning a notable portion of internet traffic, validates its capacity to enhance HTTPS traffic performance, marking a pivotal step in web protocol evolution~\cite{QUICDesignInternetScale}

A study by Yu et al.~\cite{dissecting_QUIC}, performed on production endpoints including Facebook, Google, and Cloudflare, suggests that QUIC's performance advantages, such as reduced handshake times, largely depend on the server's congestion-control decisions and client configuration specifics. Additionally, Kakhki et al.~\cite{LonglookatQUIC} analyzed multiple versions of QUIC to demonstrate that although QUIC typically surpasses TCP in performance, its effectiveness can be compromised by factors like window sizes, packet reordering, and its behavior on mobile devices and over cellular connections.

Transitioning to HTTP/3 prioritization, a study by Marx et al.~\cite{marx_2019_resource} analyzing 10 different QUIC implementations alongside 11 prioritization strategies reveals significant variations in performance. In specific scenarios, some approaches can lead to a fivefold decrease in page load times, signifying improvement. However, these benefits depend highly on the context; strategies that excel in one situation may perform poorly in another.

Additionally, Sander et al.~\cite{sander2022analyzing} indicate that while sequential scheduling suits bursty loss scenarios well, parallelism improves performance under random loss. For moderate loss, priority-aware parallelism outperforms round-robin.

Challenges in resource loading highlight the importance of practical evaluation tools to enhance website performance. Lighthouse\cite{lighthouse}, an open-source industry standard tool used in previous web performance studies\cite{saif2020early,gupta,gupta2024improving}, offers a set of metrics essential for this purpose. 

Our preceding research~\cite{gupta2024improving} delved into the incremental mechanism and discovered through experimental analysis that urgency-based incremental prioritization enhances the QoE as measured by Lighthouse. While we generally observed improvements, there were instances of slight performance degradation, underscoring the sensitivity of QoE to web page structure. 

In contrast, this paper explores the non-incremental, urgency-based resource delivery method specified by the EPS to understand its effects on QoE. This exploration seeks to uncover how such a method impacts QoE, particularly web page rendering.

\section{Proposed Mapping}
\label{proposed_mapping}

The EPS categorizes web resources based on urgency levels, ranging from 0 to 7, where 0 represents the highest urgency, and 7 is the lowest. Given the interdependence of webpage resources, assigning the correct urgency level to each resource is crucial for timely delivery. Thus, accurately mapping resources to their appropriate urgency levels is critical for optimizing the delivery sequence and enhancing the overall QoE. 

In this paper, we focus on an urgency-based, non-incremental resource delivery approach as specified by the EPS framework. We observed that Chromium does not communicate resource priorities, such as those for HTML, images, scripts, and stylesheets to the server. We implemented EPS-based urgency mapping on the server side to overcome this lack of priority communication.

To enhance the QoE, we implemented various mapping strategies. Initially, Chromium priorities were extracted from Lighthouse reports and subsequently mapped to the EPS.  This approach, termed \emph{Direct Mapping} (DM), uses one to one mapping of Chromium priorities to EPS urgency levels as detailed in Table~\ref{basic_mapping}.

\begin{figure}[p] 
  \centering
  
  \begin{subfigure}[b]{0.41\textwidth}
    \includegraphics[width=\textwidth]{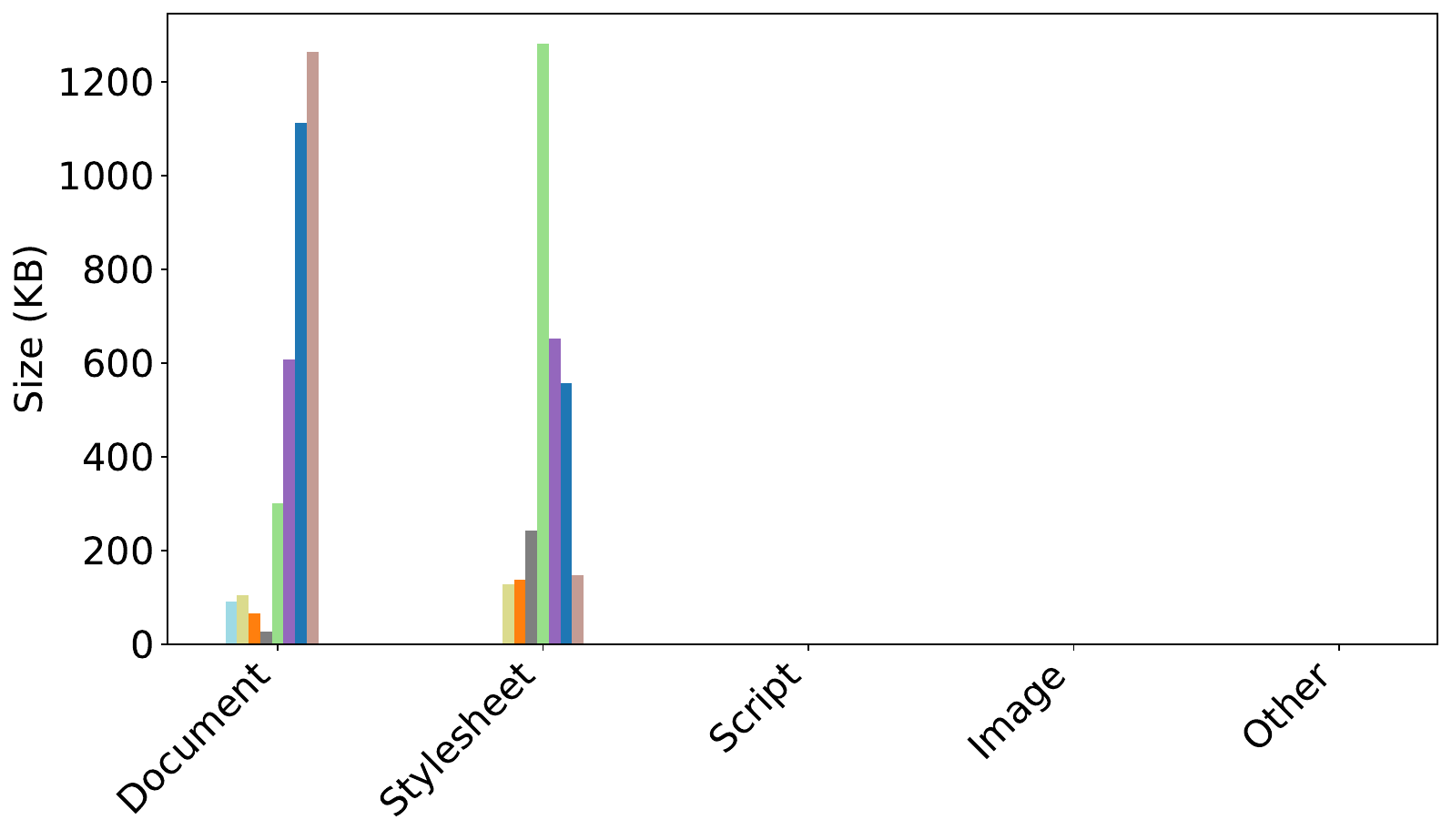}\vspace{-5pt}
    \caption{Distribution of Resources for Very High Priority}
    \vspace{15pt}
    \label{very_high_priority}
  \end{subfigure}%
  \hfill
  \begin{subfigure}[b]{0.41\textwidth}
    \includegraphics[width=\textwidth]{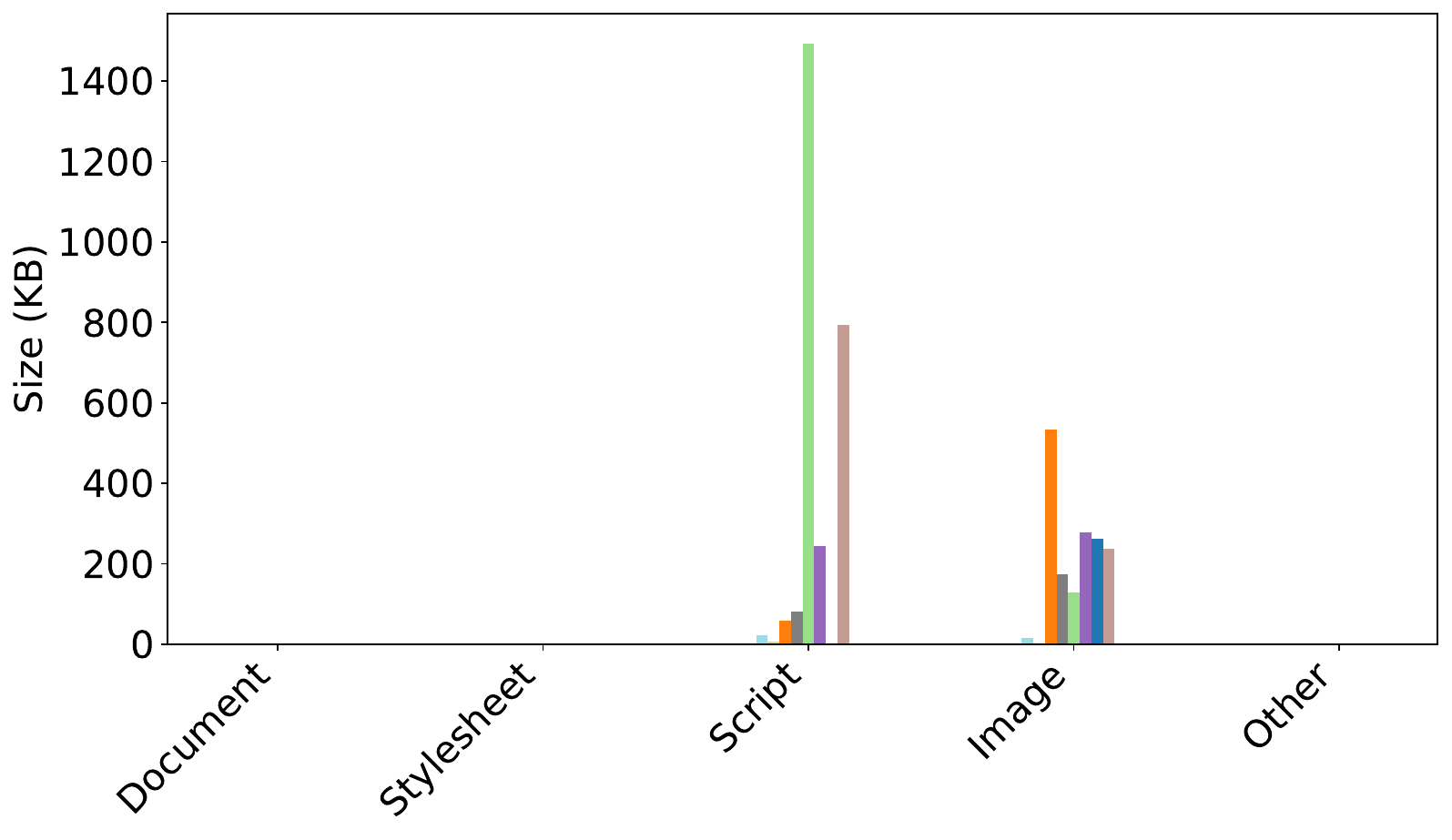}\vspace{-5pt}
    \caption{Distribution of Resources for High Priority}
    \vspace{15pt}
    \label{high_priority}
  \end{subfigure}
  
  \begin{subfigure}[b]{0.41\textwidth}
    \includegraphics[width=\textwidth]{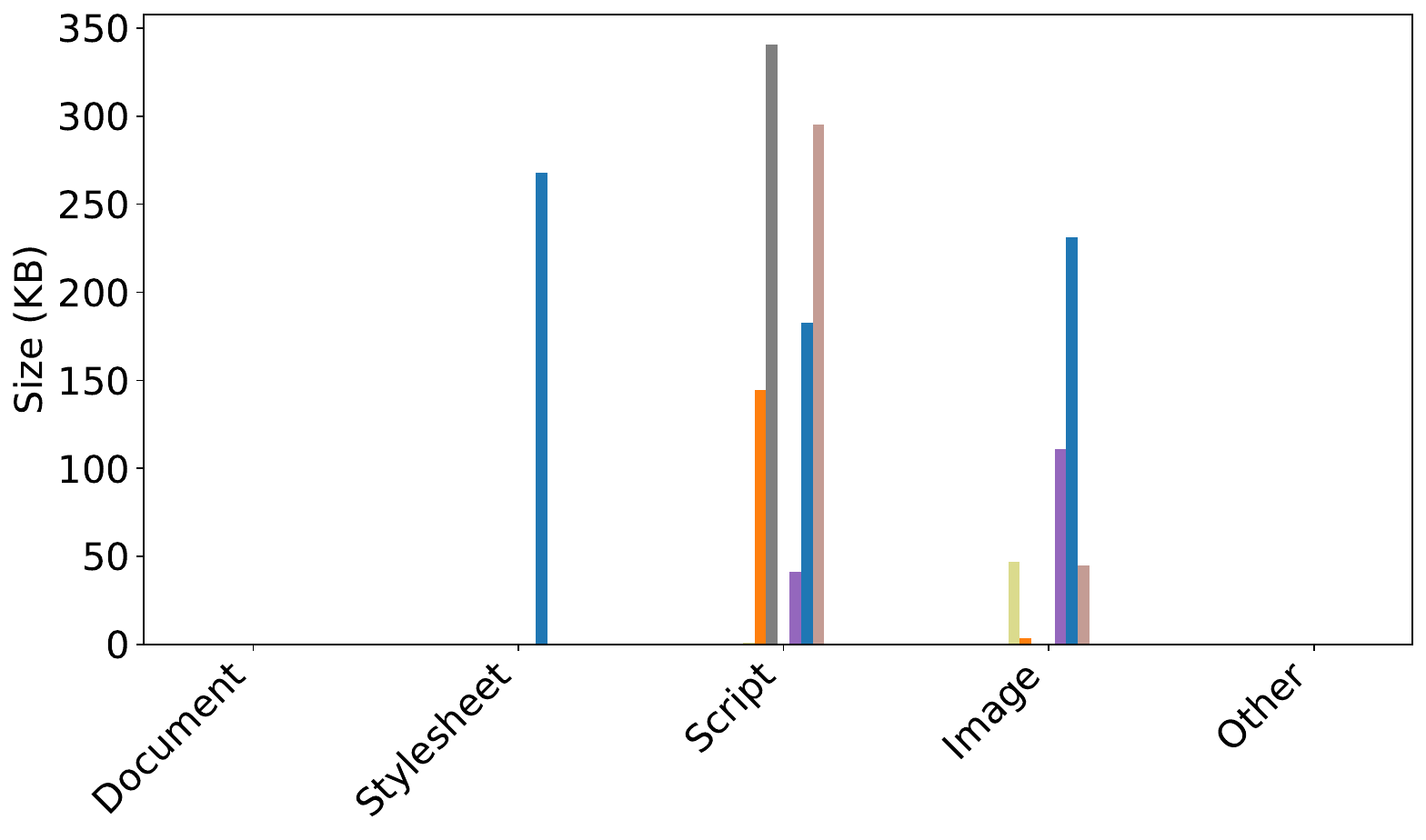}\vspace{-5pt}
    \caption{Distribution of Resources for Medium Priority}
    \vspace{15pt}
    \label{medium_priority}
  \end{subfigure}%
  \hfill
  \begin{subfigure}[b]{0.41\textwidth}
    \includegraphics[width=\textwidth]{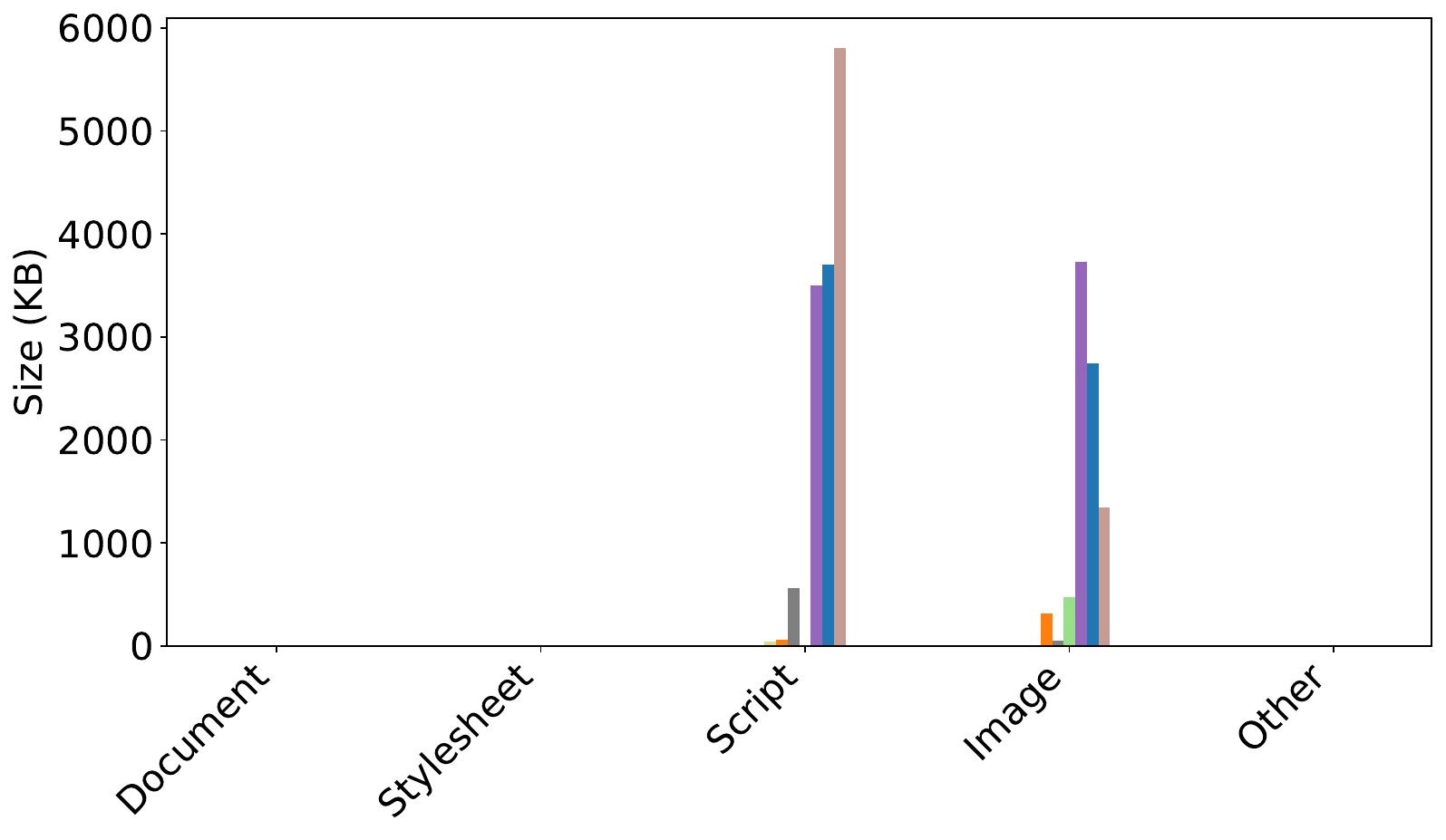}\vspace{-5pt}
    \caption{Distribution of Resources for Low Priority}
    \vspace{15pt}
    \label{low_priority}
  \end{subfigure}

 \includegraphics[width=0.40\textwidth]{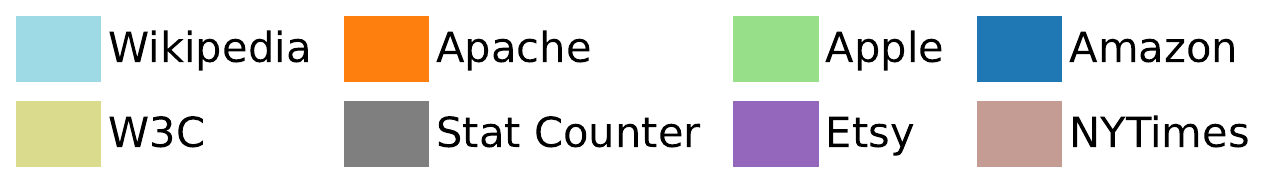}
  \vspace{12pt}
  \caption{Number of Bytes per Chromium Priority per Resource Type}
  \label{resources_by_priority_level}
 
\end{figure}

We adopted an advanced mapping strategy to enhance further the QoE, detailed in Table~\ref{enhanced_mapping}. This strategy, named \emph{Resource Type Aware Mapping} (RTAM), involved extracting resource priorities from Lighthouse reports and categorizing their types across priority levels, as shown in Figure~\ref{resources_by_priority_level}. Note that the chart for Very Low priority level was omitted because only resources in the \emph{other} category were observed, and they represented only a small fraction of the transmitted data.

\begin{figure}[ht] 
  \centering
  \begin{subfigure}[b]{0.47\textwidth}
    \includegraphics[width=\textwidth]{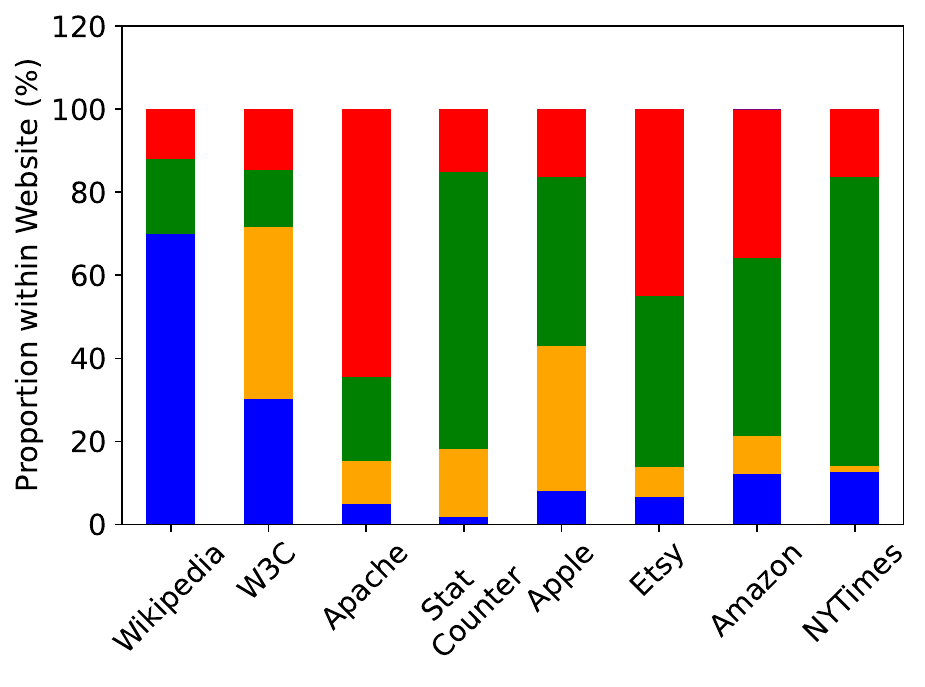}
   \vspace{-15pt}
    \caption{By Size}
    \label{plot_type_distribution_by_size}
\vspace{15pt}
  \end{subfigure}
  \hspace{0.05\textwidth} 

  \begin{subfigure}[b]{0.47\textwidth}
    \includegraphics[width=\textwidth]{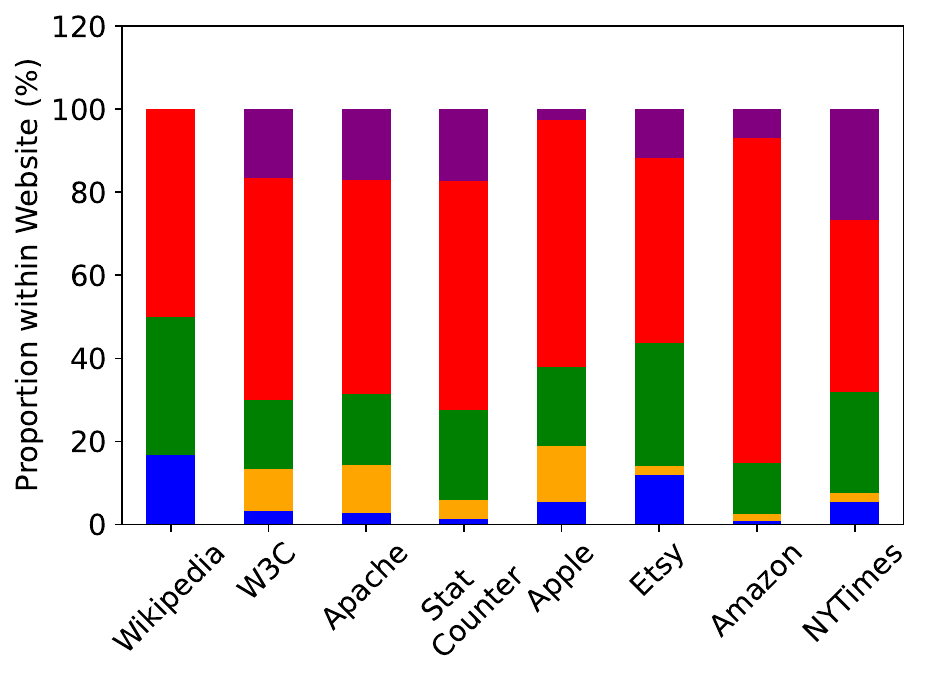}
   \vspace{-15pt}
    \caption{By Count}
     \label{plot_type_distribution_by_count} 
  \end{subfigure}

  \vspace{20pt} 

  \includegraphics[width=.48\textwidth]{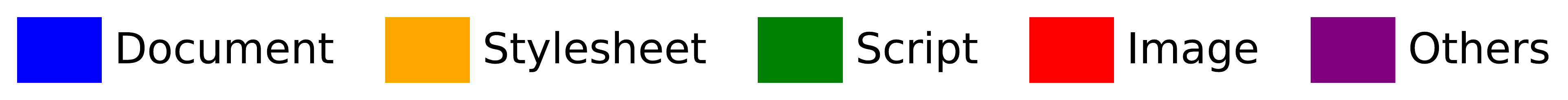}
  \vspace{10pt}
  \caption{Resources Distribution per Website}
  \label{resources_distribution_per_website}

\end{figure}

Furthermore, the resources falling into the \emph{other} category are not discernible in Figure~\ref{resources_by_priority_level} due to their small size. This pattern is corroborated by Figure~\ref{plot_type_distribution_by_size}. However, the figure categorizing resources by count, presented in Figure~\ref{plot_type_distribution_by_count}, clearly indicates the presence of these \emph{other} resources.

Our analysis of Lighthouse reports indicated a distribution pattern where scripts and images were spread from High to Low priority levels, while documents and stylesheets were invariably classified as Very High. Chromium's priority system spans five levels, ranging from Very High to Very Low, whereas EPS encompasses eight levels, offering a broader spectrum for distributing web resources. 

If a resource type is not covered in the RTAM, it defaults to the DM, with its urgency determined by Table~\ref{basic_mapping}. For example, in the medium priority level resource distribution (see Figure~\ref{medium_priority}), the stylesheet is an outlier. This deviation was assigned an urgency value of 3 due to the RTAM not specifying an urgency for this type of resource at the medium level, thereby triggering a fallback to the DM as detailed in Table~\ref{basic_mapping}. As can be seen from Figure~\ref{resources_by_priority_level}, such outliers are uncommon.

\begin{table}[tbh]
\caption{DM (Direct Mapping) from Chromium to EPS}
\label{basic_mapping}
\centering
\renewcommand{\arraystretch}{1.5} 
\begin{tabular}{|l|c|}
\hline
\multirow{2}{*}{\textbf{Chromium Priority}} & \multirow{2}{*}{\textbf{EPS Urgency}} \\ 
& \\
\hline
\textbf{Very High (0)} & 0\\ 
\hline
\textbf{High (1)} &  2  \\
\hline
\textbf{Medium (2)} & 3  \\
\hline
\textbf{Low (3)} &  5 \\
\hline
\textbf{Very Low (4)} & 7  \\
\hline
\end{tabular}
\end{table}

\begin{table}[tbh]
\caption{RTAM (Resource Type Aware Mapping) from Chromium to EPS}
\label{enhanced_mapping}
\centering
\renewcommand{\arraystretch}{1.5} 
\setlength{\tabcolsep}{4pt} 
\begin{tabular}{|l|c|c|c|c|c|}
\hline
\multicolumn{1}{|l|}{\parbox{1cm}{\vspace{4pt}\centering \textbf{Chrom.} \\ \textbf{Priority}\vspace{4pt}}} & \multicolumn{5}{c|}{\textbf{EPS Urgency}} \\
\cline{2-6}
 & \textbf{Document} & \parbox{1cm}{\vspace{4pt}\centering \textbf{Style} \\ \textbf{Sheet}\vspace{4pt}} & \textbf{Script} & \textbf{Image} & \textbf{Others} \\
\hline
\multirow{2}{*}{\parbox{1.2cm}{\vspace{4pt}\textbf{Very} \\ \textbf{High}\vspace{4pt}}} & \multirow{2}{*}{0} & \multirow{2}{*}{1} &  &  &  \\
 &  &  &  &  &  \\
\hline
 \textbf{High} &  &  & 2 & 3 &  \\
\hline
\textbf{Medium} &  &  & 4 & 5 &  \\
\hline
\textbf{Low} &  &  & 6 & 6 &  \\
\hline
\multirow{2}{*}{\parbox{1.2cm}{\vspace{4pt}\textbf{Very} \\ \textbf{Low}\vspace{4pt}}} &  &  &  &  & \multirow{2}{*}{7} \\
 &  &  &  &  &  \\
\hline
\end{tabular}
\end{table}

This expanded granularity in EPS allows for improved web resource delivery order, directly contributing to improvements in critical performance metrics such as LCP, FCP, and TTI.  

One approach in our exploration of various mapping strategies involved categorizing images and scripts into High, Medium, and Low priority levels based on their average sizes. Crucially, we preserved the integrity of resources already classified at the Very High priority level, making no alterations to their status. We determined a threshold based on the average size of images and scripts to implement this. Resources exceeding this threshold were assigned a higher urgency level, whereas those falling below it were assigned a lower urgency. Despite the rationale behind this strategy, it unfortunately failed to produce the anticipated improvements in performance.

\begin{figure}[t]
  \includegraphics[width=\linewidth]{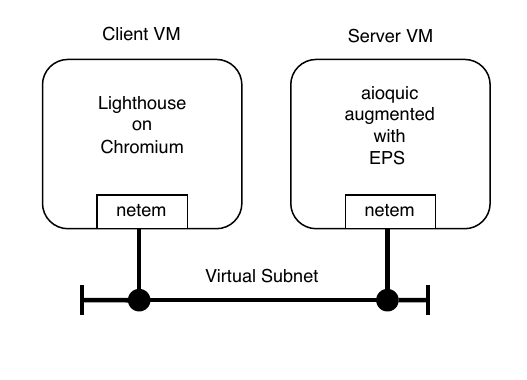}
  \vspace{-35pt}
  \caption{Experimental Setup}
  \label{Experimental_Setup}
\end{figure}

\section{Experimental Setup}
\label{experimental_setup}

This paper defines the experimental framework using two virtual machines (VMs), the Client VM and the Server VM, which are interconnected through a virtual subnet to create a controlled testing environment. These VMs are hosted on a MacBook Pro that features macOS Sonoma version 14.3.1, 64 GB of RAM, and an M1 CPU, with Parallels Desktop 19~\cite{parallels} for MAC employed for virtualization. Each VM runs Ubuntu 22.04.2 with a Linux 5.15.0-76-generic kernel and is provisioned with 16 GB of RAM and 64 GB of storage. Experiments were conducted in isolation without internet connectivity to eliminate external influences.

The Client VM hosts Lighthouse version 11.2.0 on Chromium version 118.0.5993.70 to perform audits on websites' performance, simulating the end-user experience. Meanwhile, the Server VM hosts two distinct versions of the \emph{aioquic} server: the original out-of-the-box version and a modified version incorporating an urgency-based non-incremental resource delivery method specified by the EPS. This augmentation with our prioritization mapping strategies like DM and RTAM improves the QoE, making it the focal point of our experimental evaluation.

\begin{figure}[t]
  \includegraphics[width=\linewidth]{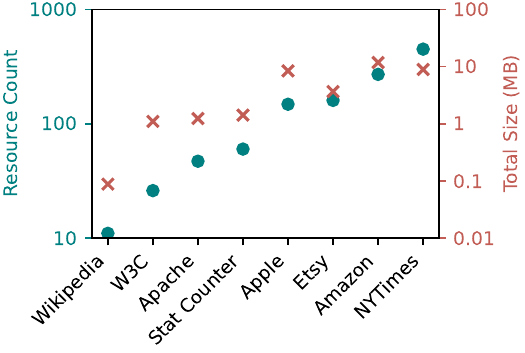}
  \caption{Resource Counts and Total Sizes}
  \label{websites_with_resources}
\end{figure}

\emph{Netem}~\cite{NetEm} was used for emulating realistic network traffic conditions. 10 ms latency and 0.05 \% loss were employed on the network interface of each virtual machine in each direction. 

Eight popular websites, shown in Figure~\ref{websites_with_resources} were downloaded. The websites are sorted in the figure from lowest to the highest resource counts, with Wikipedia having the lowest number of resources and the New York Times having the highest. 

External trackers were eliminated from several websites to maintain a controlled testing environment. This step was taken to avoid downloading extraneous content and preserve the integrity of our testing configuration. Each website was subjected to ten audits to ensure accuracy and minimize potential errors.

\begin{figure}[t]
  \includegraphics[width=\linewidth]{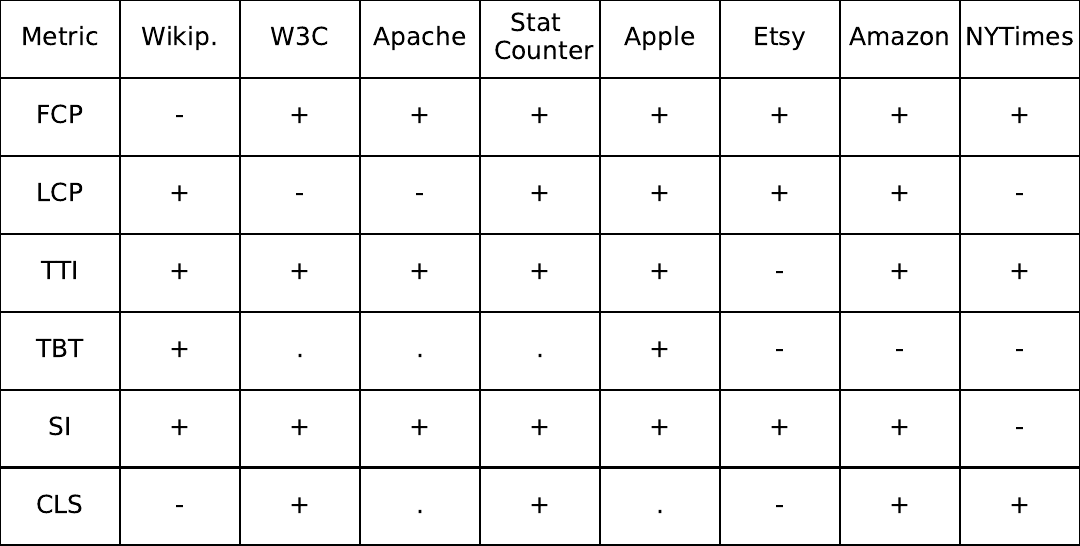}
  \caption{Relative Lighthouse Metrics Improvements of RTAM over Sequential Delivery}
  \label{pattern}
\end{figure}

\begin{figure}[t]
  \includegraphics[width=\linewidth]{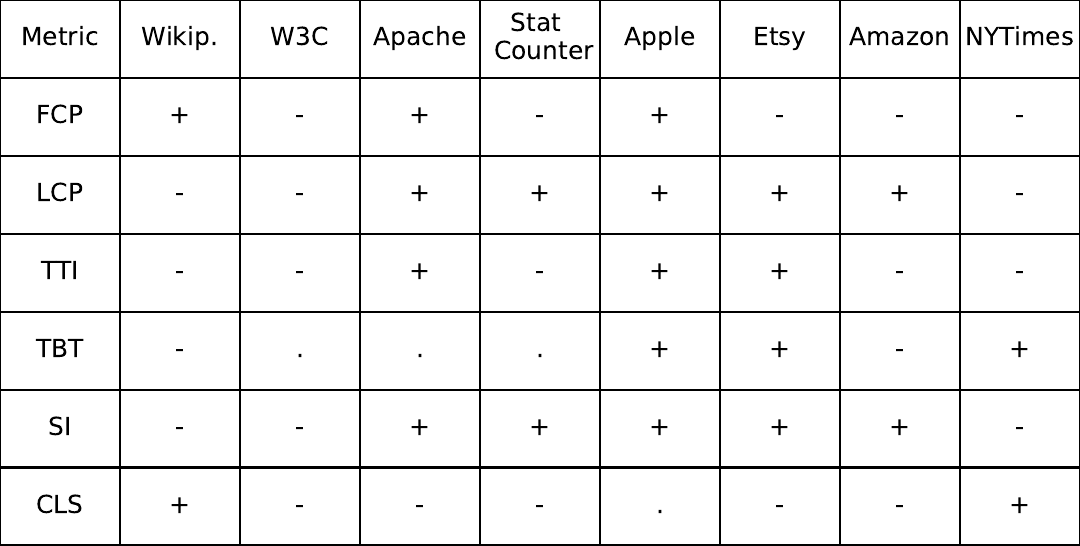}
  \caption{Relative Lighthouse Metrics Improvements of RTAM over DM}
  \label{RTAM_vs_DM}
\end{figure}

\begin{figure*}[htbp!]
  \centering
  \newlength{\figheight}
  \setlength{\figheight}{.18\textheight} 
  \begin{subfigure}[b]{0.44\textwidth}
    \includegraphics[width=\textwidth, height=\figheight, keepaspectratio]{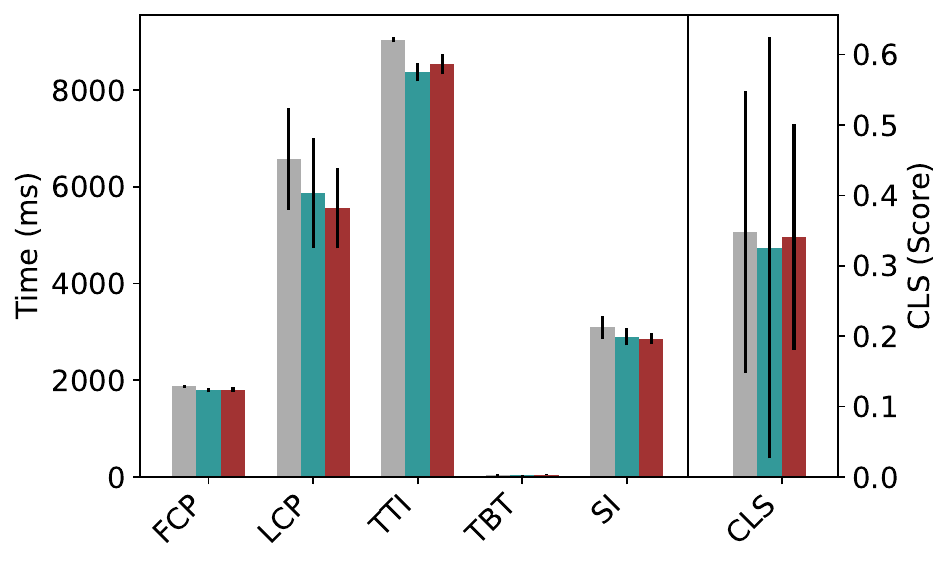}
    \caption{www.amazon.com}
    \label{amazon}
  \end{subfigure}
  \hspace{0.02\textwidth} 
  \begin{subfigure}[b]{0.44\textwidth}
    \includegraphics[width=\textwidth, height=\figheight, keepaspectratio]{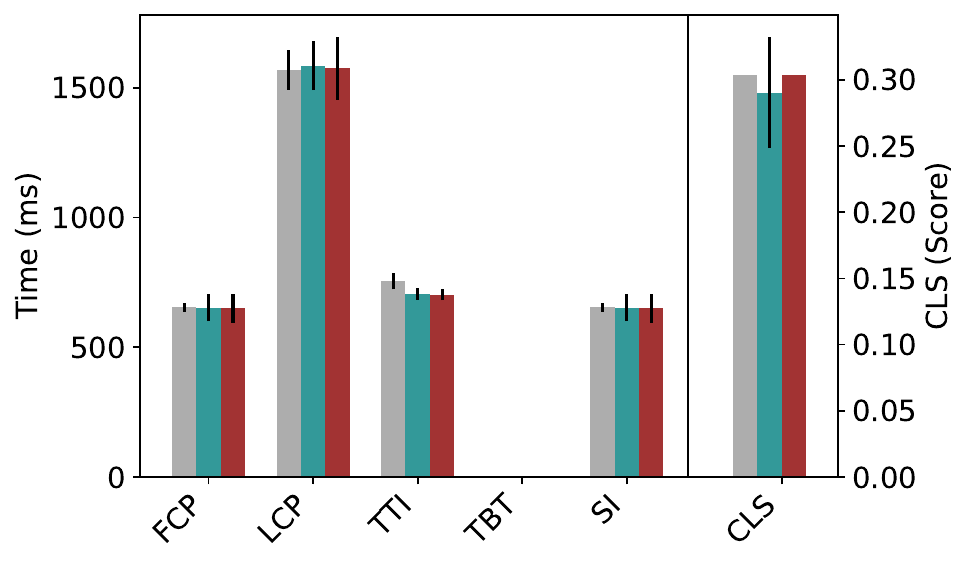}
    \caption{www.apache.com}
    \label{apache}
  \end{subfigure}
  \newline

  \begin{subfigure}[b]{0.44\textwidth}
    \includegraphics[width=\textwidth, height=\figheight, keepaspectratio]{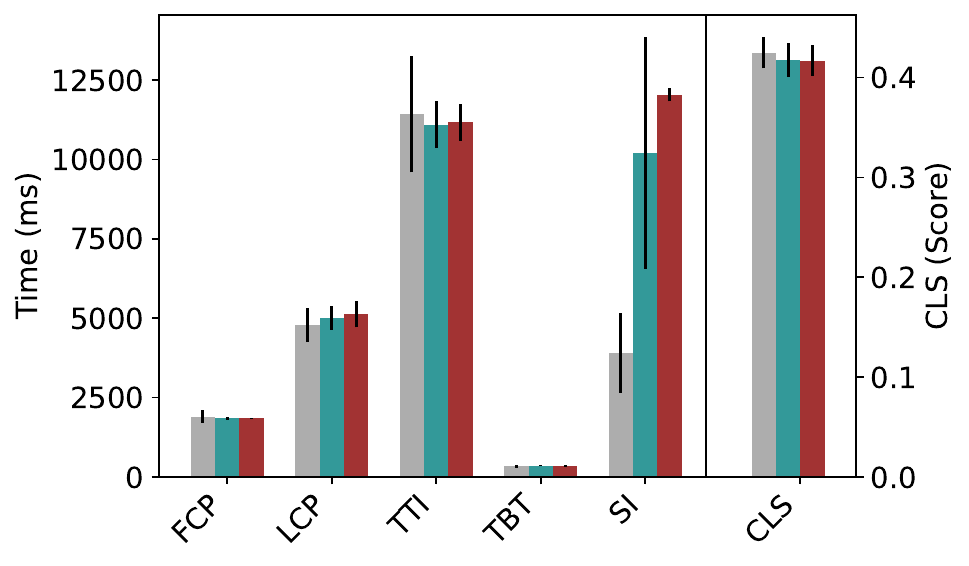}
    \caption{www.nytimes.com}
    \label{nytimes}
  \end{subfigure}
  \hspace{0.02\textwidth} 
  \begin{subfigure}[b]{0.44\textwidth}
    \includegraphics[width=\textwidth, height=\figheight, keepaspectratio]{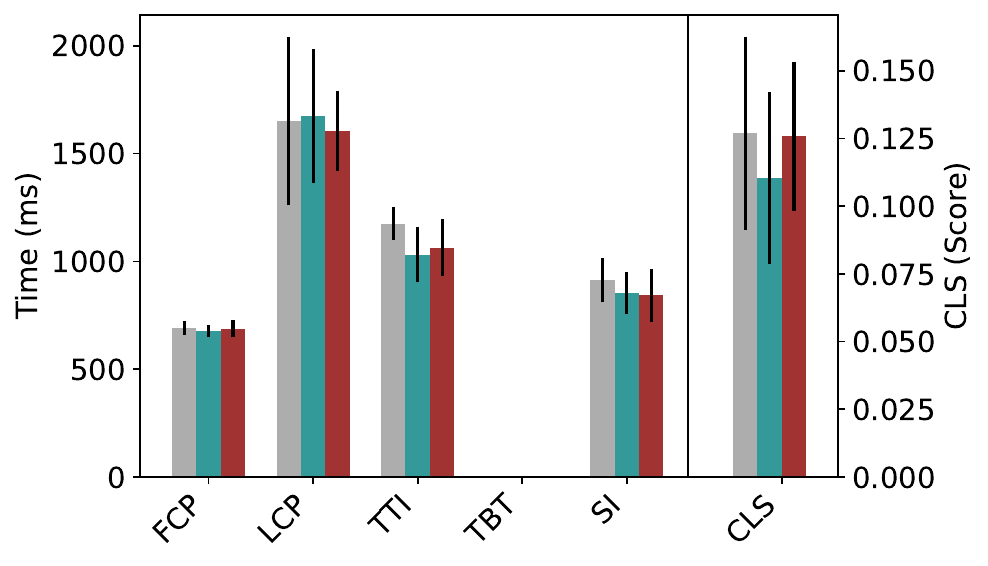}
    \caption{www.statcounter.com}
    \label{statcounter}
  \end{subfigure}
  \newline

  \begin{subfigure}[b]{0.44\textwidth}
    \includegraphics[width=\textwidth, height=\figheight, keepaspectratio]{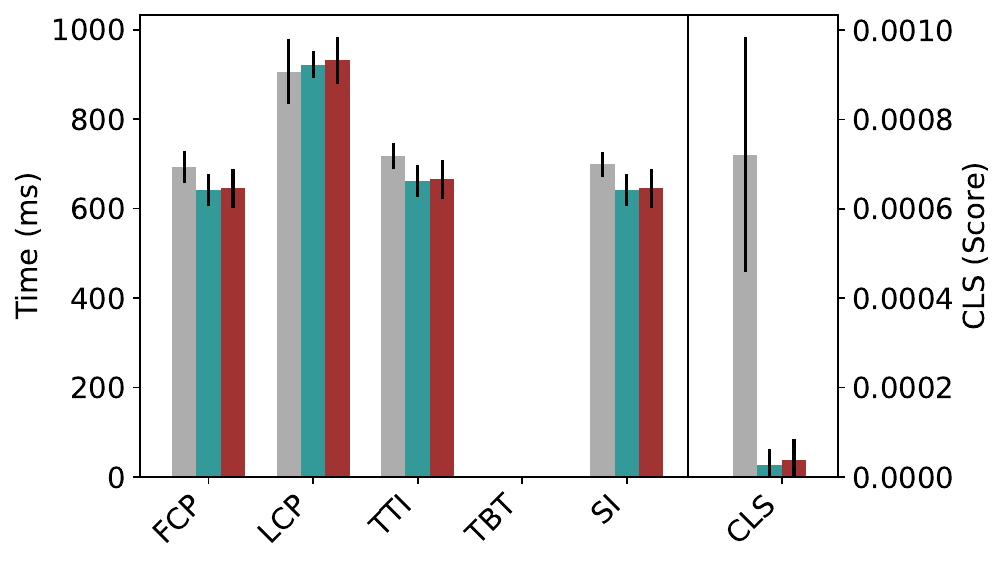}
    \caption{www.w3.org}
    \label{w3c}
  \end{subfigure}
  \hspace{0.02\textwidth} 
  \begin{subfigure}[b]{0.45\textwidth}
    \includegraphics[width=\textwidth, height=\figheight, keepaspectratio]{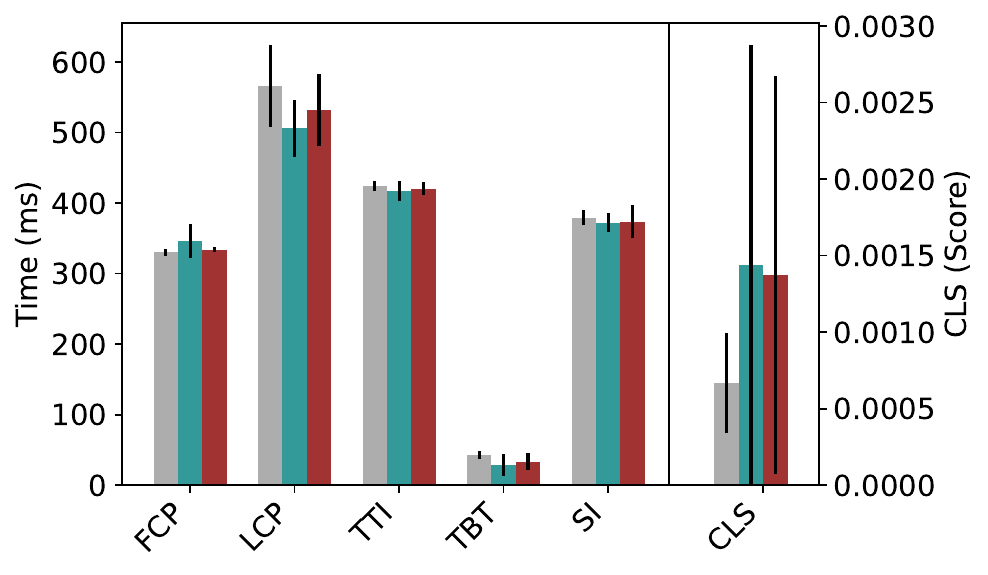}
    \caption{www.wikipedia.org}
    \label{wikipedia}
  \end{subfigure}
  \newline

  \begin{subfigure}[b]{0.44\textwidth}
    \includegraphics[width=\textwidth, height=\figheight, keepaspectratio]{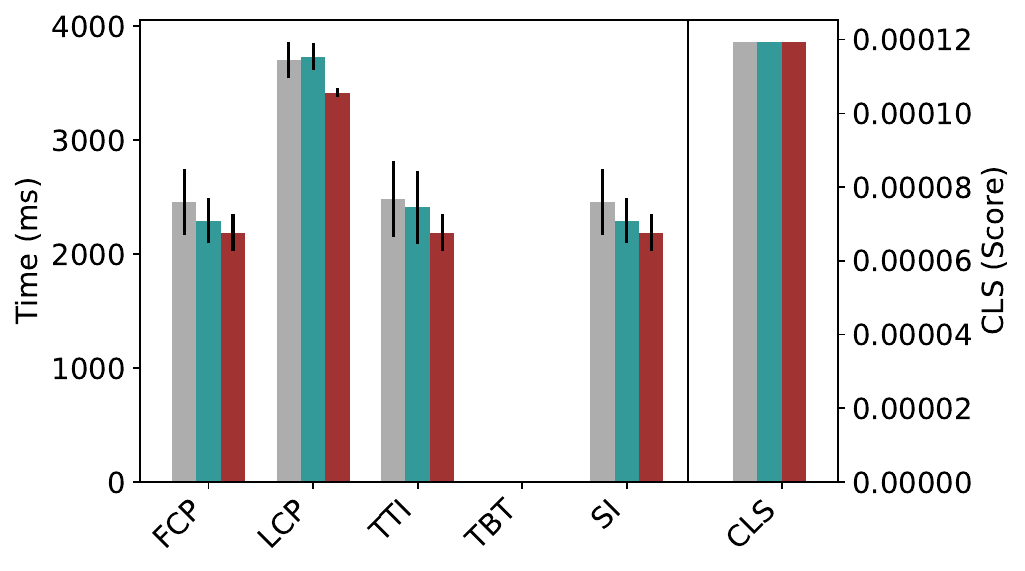}
    \caption{www.apple.com}
    \label{apple}
  \end{subfigure}
  \hspace{0.02\textwidth} 
  \begin{subfigure}[b]{0.44\textwidth}
    \includegraphics[width=\textwidth, height=\figheight, keepaspectratio]{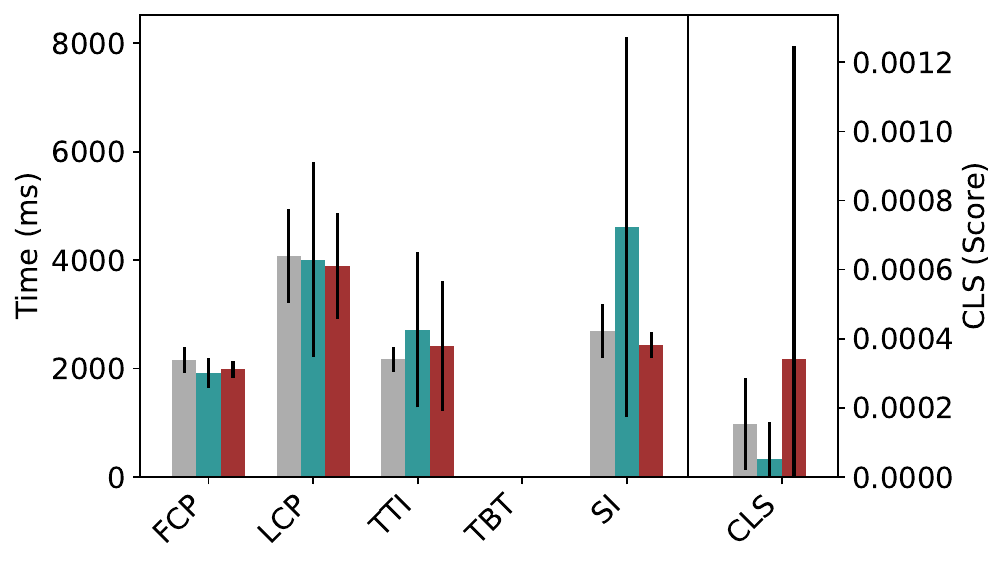}
    \caption{www.etsy.com}
    \label{etsy}
  \end{subfigure}
  \newline

  \includegraphics[width=0.7\textwidth]{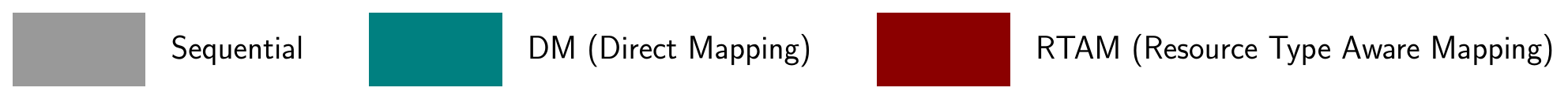} 

  \caption{Comparison of Web Performance Metrics across different Websites.}
  \label{barchart}
\end{figure*}

\section{Experimental Evaluation}
\label{experimental_evaluation}
In this section, we present our study's results, contrasting the original \emph{aioquic} implementation, which uses a sequential (non-incremental) resource delivery method, against our augmented \emph{aioquic} that adopts the urgency-based non-incremental prioritization framework as specified by the EPS and employs both DM and RTAM, showcasing performance improvements. 

In Figure~\ref{barchart}, three values are shown for each metric. The series initiates with default \emph{aioquic}'s sequential delivery, augmented \emph{aioquic} with an urgency-based, non-incremental resource delivery method alongside DM and \emph{aioquic} combined with RTAM. Figure~\ref{barchart} uses dual scales to accommodate different types of metrics. On one side, metrics such as FCP, LCP, TTI, TBT, and SI are presented, all measured in time units. On the opposite side, CLS is displayed, a unitless value derived from the extent and magnitude of content shifts on the page. A vertical line in Figure~\ref{barchart} separates CLS from the time-based metrics.

Moreover, Figure~\ref{pattern} offers a summary of the performance improvements achieved across the websites and metrics. Positive signs indicate areas where the urgency-based non-incremental resource delivery method with RTAM has surpassed the performance of the sequential strategy. Negative signs, conversely, indicate metrics or websites where it underperformed. Dots within the matrix denote instances of no performance change between the two methods. The relative improvement of RTAM over DM is summarized in Figure~\ref{RTAM_vs_DM}.

\subsection{First Contentful Paint and Largest Contentful Paint}

Improvements in FCP scores with RTAM compared to sequential delivery are demonstrated, with most websites experiencing improvements, as shown in Figure~\ref{pattern}. The FCP performance reduction observed for Wikipedia is marginal, indicating the robustness of the improvement across different sites. For Wikipedia, the FCP time increased from 330 ms to 334 ms, marking a minimal decrease in performance. Nonetheless, this adjustment remains within the good range of FCP thresholds as defined by Chrome Developers~\cite{lighthouse_performance_scoring}.

LCP scores also saw improvements across most evaluated websites. A slight performance dip in LCP for Apache, alongside slightly more pronounced degradations for New York Times and W3C, can be attributed to the prioritization mapping schema employed. Furthermore, it is worth noting that DM worked better for Wikipedia, suggesting the influence of website-specific factors such as small number of resources and small total size on performance outcomes.

In the case of the New York Times, the number and size of scripts surpass those of images. Our DM and RTAM prioritization strategies assigned greater EPS urgency levels to images to enhance the LCP scores. However, this approach faced challenges, given the intricate architecture and the closely integrated nature of scripts and images. As a result, the delivery of images critical to LCP was postponed, impacting the LCP performance. 

For both W3C and Apache, there was a minor reduction in the LCP metric. Nevertheless, this reduction maintains the LCP within the good range, of 0 to 2.5 seconds as specified by Chrome Developers~\cite{lighthouse_performance_scoring}. This indicates that, according to this benchmark, the performance continues to be satisfactory.

\subsection{Time To Interactive and Total Blocking Time}

TTI experienced improvements across most of the websites analyzed, with Etsy being an exception, as illustrated in Figure~\ref{pattern}. Conversely, Total Blocking Time (TBT) saw improvements for Apple and Wikipedia, while it underperformed for Etsy, Amazon, and the New York Times.

The performance metrics for individual websites under all three resource delivery methods, highlighting a marginal decline in TTI performance for Etsy, are detailed in Figure~\ref{barchart}. This reduction stems from adopting an urgency-based non-incremental resource delivery method, which blocks the delivery of lower-urgency resources until the higher-urgency ones have been delivered. In Etsy's case, the intentional delay in script delivery was a strategic decision aimed at enhancing other critical performance metrics, namely FCP and LCP, at the cost of prolonged interactivity readiness. 

Among RTAM and DM, we observe mixed results. For Apache, Apple, and Etsy, RTAM outperformed DM, while it marginally deteriorated in other cases. Although this approach led to a decrease in performance compared to sequential delivery, it is essential to note that, according to Chrome Developers~\cite{lighthouse_performance_scoring}, any TTI score below 3.8 seconds is considered good. The TTI score achieved through our urgency-based non-incremental method remains well under this threshold.

Furthermore, a marginal decrease in TBT performance was observed for Etsy, Amazon, and the New York Times. This reduction in performance originates from our non-incremental delivery method, which blocks scripts to deliver other resources. It is crucial to note that, according to Chrome Developers~\cite{lighthouse_performance_scoring},any TBT score below 200 ms is considered good. In our evaluation, both Etsy and Amazon achieved TBT scores well under this threshold, showcasing the effectiveness of our approach.

\subsection{Speed Index}

SI experienced improvements across the websites, except for the New York Times, as illustrated in Figure~\ref{pattern}. This improvement indicates that our urgency-based non-incremental approach with mapping strategies for delivering web resources effectively accelerated the delivery and rendering of crucial page components.

SI saw improvements for all analyzed websites, apart from the New York Times, as detailed in Figure~\ref{barchart}. The latter's performance lagged primarily due to its higher proportion of scripts relative to images. In efforts to improve FCP, scripts were deprioritized and assigned lower urgency, leading to a diminished speed index for the New York Times. Furthermore, it is worth noting that both mapping strategies decreased performance for the New York Times, with RTAM showing a more pronounced decline.

Comparing DM and RTAM, SI showed better performance when used with RTAM on most websites. However, an exception was observed with the New York Times, where there was a decline in SI's performance. For Wikipedia and W3C, the performance results were consistent between DM and RTAM.

\subsection{Cumulative Layout Shift}

An improvement in CLS for four websites when comparing RTAM to Sequential delivery, deterioration is observed for two websites, while no change is noted for the remaining two as illustrated in Figure~\ref{pattern}. The sites experiencing performance degradation are Wikipedia and Etsy.

Detailed CLS scores, showcasing the impact of our resource delivery methods, are provided in Figure~\ref{barchart}. Notably, for Etsy, the CLS value decreased to $5.13 \times 10^{-5}$
 with DM from $1.54 \times 10^{-4}$
 under sequential delivery, and increased to $3.41 \times 10^{-4}$
 with RTAM. It is worth mentioning that CLS scores for W3C, Apache, Stat Counter, Etsy, and Amazon are better under DM. 

In the case of Wikipedia, CLS values grew from $6.66 \times 10^{-4}$
in the sequential approach to $1.37 \times 10^{-3}$ under RTAM. This adjustment to boost metrics such as LCP, TTI, TBT, and SI resulted in a higher CLS. Despite this increase,  Wikipedia's and Etsy's CLS scores remain within the good range, following the Chrome Developers guidelines~\cite{lighthouse_performance_scoring}.

\subsection{Performance under more challenging network conditions}

We conducted the same set of experiments to evaluate performance under more challenging network conditions. We introduced a 0.1\% loss and a 20 ms latency on both the Client VM and the Server VM. The observed results were consistent with those described above. Due to space constraints, these findings are not detailed here.

\section{Conclusions}
\label{conclusions}

HTTP/3 leverages QUIC's stream multiplexing capability for faster data transmission. This stream multiplexing allows sequential (non-incremental) or concurrent (incremental) data delivery. This study investigates the influence of employing urgency-based
non-incremental resource delivery method as outlined in RFC 9218. We propose and apply prioritization mappings to evaluate their impact on the QoE across different websites.

The effectiveness of performance improvement is influenced by prioritization mapping. This study examines the priority levels assigned by Chromium to various resources, spanning five levels from Very High to Very Low. Leveraging insights from this analysis, we propose prioritization mapping schemes termed DM and RTAM that assign urgency levels to different resources. Subsequently, we assess the impact of our mapping scheme on website performance through the use of Lighthouse.

Experimental evaluation indicates that using both mapping schemes for urgency-based non-incremental resource delivery methods enhances the performance of metrics such as FCP, LCP, TTI, and SI metrics on most websites tested, as evaluated by Lighthouse. Moreover, we observed that the RTAM further improved LCP and SI.

As we outline future research directions, browsers are expected to provide support for EPS, which will provide a framework for designing methods to further improve web content delivery. Additionally, enabling servers to dynamically adjust client-specified priorities to dynamically optimize resource delivery for improved performance represents a significant area of exploration.

\bibliographystyle{ieeetr}
\bibliography{bibliography.bib}
\end{document}